\documentclass[preprint,12pt,3p]{elsarticle}

\usepackage{ulem}

\usepackage{amssymb}

\usepackage[usenames,dvipsnames]{xcolor}

\usepackage{gensymb}

\begin{document}

\begin{frontmatter}

\title{Correlated emission of X-ray and sound from water film irradiated by femtosecond laser pulses}

\author[label1]{Hsin-hui Huang}
\address[label1]{Research Center for Applied Sciences, Academia Sinica, Taipei 115, Taiwan}
\ead{hsinhuih@gate.sinica.edu.tw}

\author[label2]{Saulius Juodkazis\fnref{label3}\corref{cor1}}
\address[label2]{Nanotechnology Facility, Center for Micro-Photonics, Swinburne University of Technology, Hawthorn, VIC~3122, Australia}
\address[label3]{Melbourne Centre for Nanofabrication, the Victorian Node of the Australian National Fabrication Facility, Clayton, VIC 3168, Australia}
\ead{sjuodkazis@swin.edu.au}

\author[label1]{Koji Hatanaka\fnref{label4,label5}\corref{cor1}}
\address[label4]{College of Engineering, Chang Gung University, Taoyuan 33302, Taiwan}
\address[label5]{Department of Materials Science and Engineering, National Dong-Hwa University, Hualien 97401, Taiwan}

\cortext[cor1]{corresponding authors}

\ead{kojihtnk@gate.sinica.edu.tw}

\begin{abstract}
Simultaneous measurements of hard X-ray by a Geiger counter and audible sound (10 Hz-20 kHz) by a microphone from a thin water film in air were carried out under intense single and double pulse irradiations of femtosecond laser (35 fs, 800 nm, 1 kHz). Emission profiles of X-ray and sound under the single pulse irradiation by changing the water film position along the laser incident direction (Z-axis) show the same peak positions with a broader emission in sound (403~$\mu$m at FWHM) than in X-ray (37~$\mu$m). Under the double pulse irradiation condition with the time delay at 0 ps and 4.6 ns, it was clearly observed that the acoustic signal intensity is enhanced in associated with X-ray intensity enhancements. The enhancements can be assigned to laser ablation dynamics such as pre-plasma formation and transient surface roughness formation induced by the pre-pulse irradiation. For the acoustic signal under the double-pulse irradiation with the time delay, there was a weak dependence observed on the pre-pulse irradiation position at the laser focus. It is consistent with a long breakdown filament formation which makes the microphone-detection less position-sensitive.  
\end{abstract}


\end{frontmatter}


\vskip 0.8in
\section{Introduction}
\label{sec1}
Intense laser irradiation to solid materials or solutions, e.g., water, results in photon emission from X-ray (keV), visible light (eV), and down to THz radiation (meV). This phenomenon is, usually through a plasma formation~\cite{He2015}, associated with sound/ultrasound emission and macroscopic morphological changes as a result of laser ablation~\cite{ZHANG2014, Mustafa18} - removal of irradiated material. Each of the listed emissions has been utilized for applications in basic science and for industrial/medical usages~\cite{Wolbarst06, Inoue2016, nanoimg2010, mittleman2018, Roland2008, Mittleman17}. Such multi-color emission in the wide range of spectrum can be expected only under intense femtosecond (fs) laser irradiation conditions. If a fs-laser pulse is tightly focused into a spot of 10~$\mu$m diameter, when the pulse width, the wavelength, and the intensity is 35~fs (about 10~$\mu$m in spatial length), 800~nm, 1~mJ/pulse, respectively as a usual case, the laser volumetric power density in the cylindrical focal volume can be estimated to be about $3.6\times 10^{16}$~W/cm$^3$ under an ideal focusing condition. This laser power density can be re-calculated to $8.5\times 10^3$~mol/L as photon density in the volume (mol is the Avogadro number of photons), which can be comparable with the density of water molecules in solution phase at $5.5\times 10^1$~mol/L, which is more than 150 photons per one water molecule.

Water is one of the typical transparent materials and has been a target in studies of white light continuum generation~\cite{Brodeur99}, soft/hard X-ray emission~\cite{softxraybook, hatanaka2012xray, Masim2016}, sound/ultrasound emission~\cite{sound16, Wangbook}. Compared with other transparent solid materials such as glasses, a flowing water film refreshes the light-matter interaction volume, therefore producing a durable target for intense laser irradiation. Furthermore, as a polar solvent, addition of various solutes such as electrolytes (ions), metal nano-particles such as gold, and hydrogen peroxide can easily change and control light interaction with water. Recently, water was also used as a source for THz wave emission under intense laser irradiation conditions~\cite{17apl71103,NatCommun2014} though it has been considered to be non-applicable partly because of the high absorption coefficient of water 220~cm$^{-1}$~\cite{Thrane1995, Thrane1997} in the THz wave region. 

While independent emission of X-ray or THz wave has been widely reported, a simultaneous emission of X-ray and THz wave in conventional laboratories is highly expected for future applications~\cite{NatPhoton2017}. However, such reports on simultaneous emission of X-ray and THz wave are quite limited to a few reports on targets like Al-coated glass substrate~\cite{Hamster1993}, He gas~\cite{Hamster1993}, Ar gas~\cite{IEEE2016}, or water in solution phase~\cite{Hatanaka2018THz}. On the other hand, sound/ultrasound emission, as another emission from laser-induced plasma, has recently emerged as one promising source for microscopy with super-resolution~\cite{Chaigne2016, Chaigne2017, katz2017} or has been applied to an advanced analytical method for trace detection of gases at ppm level~\cite{Xiong7246}. Following these progresses in sound technology, combination usages of sound/ultrasound with other emission like X-ray or THz wave are also expected.

In this paper, hard X-ray and audible sound (10-20~kHz) emission from thin water film irradiated by focused fs-laser pulses in air is simultaneously measured under single and double pulse irradiation conditions. The correlation of the two emissions are discussed from the viewpoint of tight laser focusing and plasma formation on the water film. 

\label{subsec1}
\section{Experimental}
\label{sec2}

\begin{figure}[tb]
\centering
 \includegraphics[width=14cm]{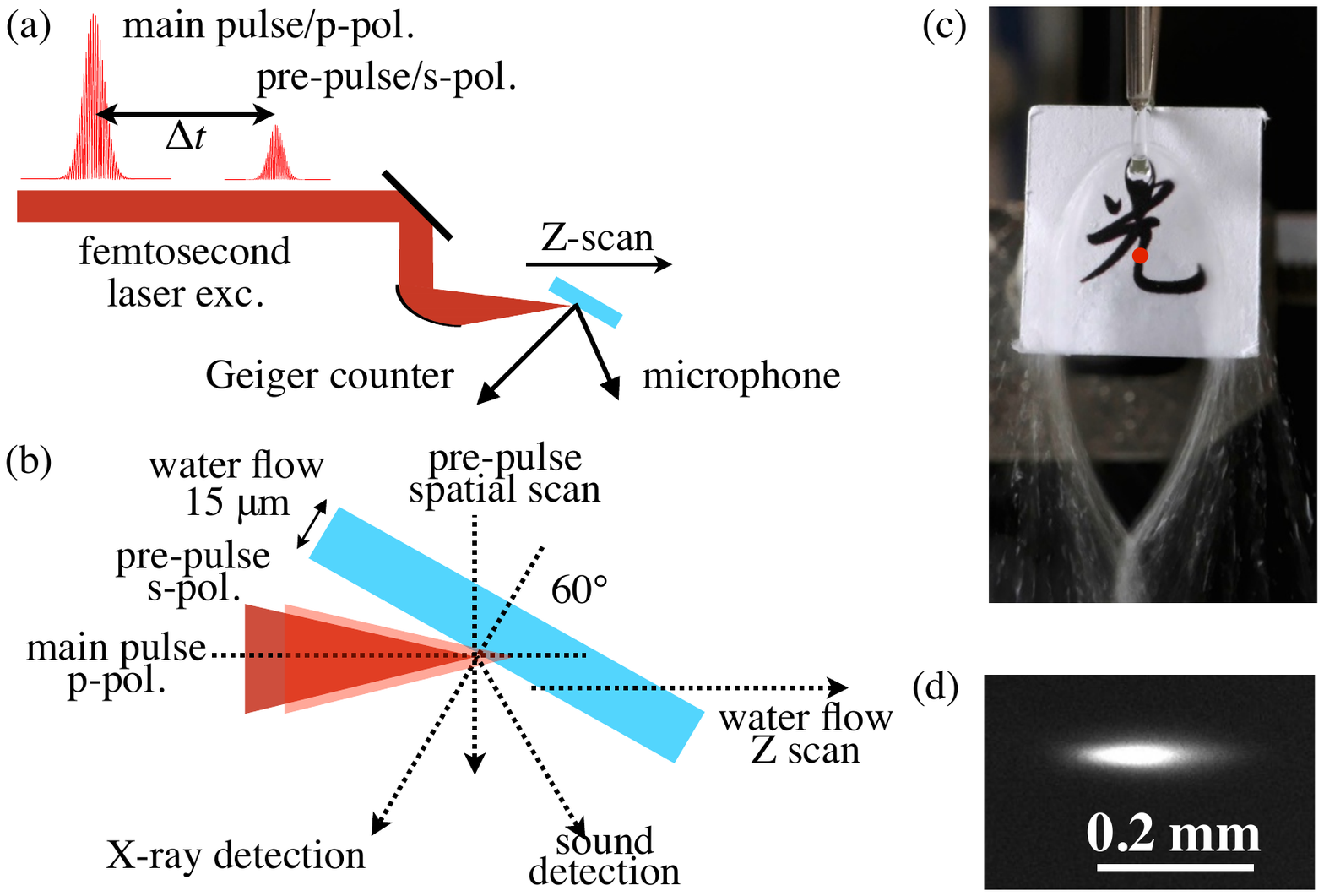}
 \caption{(a) The experimental setup (top view) for the measurements of X-ray and audible sound by changing the water flim position (Z-scan). (b) The pre-pulse irradiation position (spatial scan) under the single and the double pulse excitation conditions. (c) The water flim was made by the two colliding jets. The red dot represents the position where the incident laser pulses irradiated. (d) A photo of the air plasma observed from the side when the incident laser intensity is at 0.4 mJ/pulse.}
 \label{fgr:fig1}
\end{figure}
The experimental setup is shown in Fig.~1. Fs-laser pulses ($>$35~fs, 800~nm, linearly-polarized (horizontally), 1~kHz, Mantis/Legend Elite HE USP, Coherent, Inc.) were tightly focused in air by an off-axis parabolic mirror (the effective focus length of 50.8~mm, the reflection angle of 90\degree, and the numerical aperture $NA =0.25$, 47-097, Edmund Optics) to a water film prepared by a metal nozzle (Flatjet Nozzle LARGE, Metaheuristic). Under these focusing conditions, the theoretical diameter of the focal spot is $2w_0 = 1.22\lambda/NA = 3.9~\mu$m and the axial extent along the propagation estimated as two Rayleigh lengths $2z_R =2n\frac{\lambda}{NA^2} = 25.5~\mu$m where $n = 1$ is refractive index of air. The water film thickness was estimated to be about 15~$\mu$m by a laser-based thickness measurement (SI-T80, Keyence). The nozzle was set with a rotation and automatic-3D stages (KS701-20LMS, Suruga Seiki) and the solution angle and the position was finely optimized. The water film position along the laser incident direction (z-axis) was measured by a laser displacement sensor (LK-G80, Keyence). The laser incident angle was set at 60\degree~to the water film normal. Under this experimental setup, the spatial resolution of the water film along the Z-axis is limited by the automatic stage at 1~$\mu$m and precisely confirmed by the displacement sensor. The water film rate was estimated to be about 70 mL/min, therefore fresh and smooth water film surface is prepared for each laser irradiation every 1~ms at 1~kHz repetition rate. 

X-ray intensity was measured by a Geiger counter (SS315, Radhound, Southern Scientific) set at 11 cm distance from the laser focus with an appropriate iris on its head. The sound emission was measured in the frequency region between 10 Hz and 20 kHz by a microphone (4158N, Aco Co., Ltd.) set at 3 cm distance from the laser focus at 60\degree from the water film normal. The background level of the sound intensity only from air-plasma without the water film detected under this condition is negligibly small and the background level has been subtracted in advance from all the sound data shown in the figures. The temporal response function of the Geiger counter is slow at 1 sec. while the function of the microphone is faster at 20 ~$\mu$s under the repetitive laser irradiation at 1 kHz. A combination of a half-wave plate (65-906, Edmund Optics) and a polarization-sensitive beam splitter (47-048, Edmund Optics) divides the laser beam into two different beams, the main pulse (0.4 mJ/pulse, horizontally-pol./p-pol. to the water film surface) and a pre-pulse (0.1 mJ/pulse, s-pol.). Double pulse excitation with time delay was carried out with and an automatic optical delay stage (SGSP46-800, Sigma Koki). Experiments on spatial (horizontal) scan of the pre-pulse at the laser focus on the water film was carried out with an automatic scanning mirror (POLARIS-K2S2P, Thorlabs) with a piezo controller (KPZ101, Thorlabs). All experiments were performed at room temperature (296 K) and at ambient air pressure.

\section{Results and discussion}\label{sec3}
\begin{figure}[tb]
\centering
 \includegraphics[width=14cm]{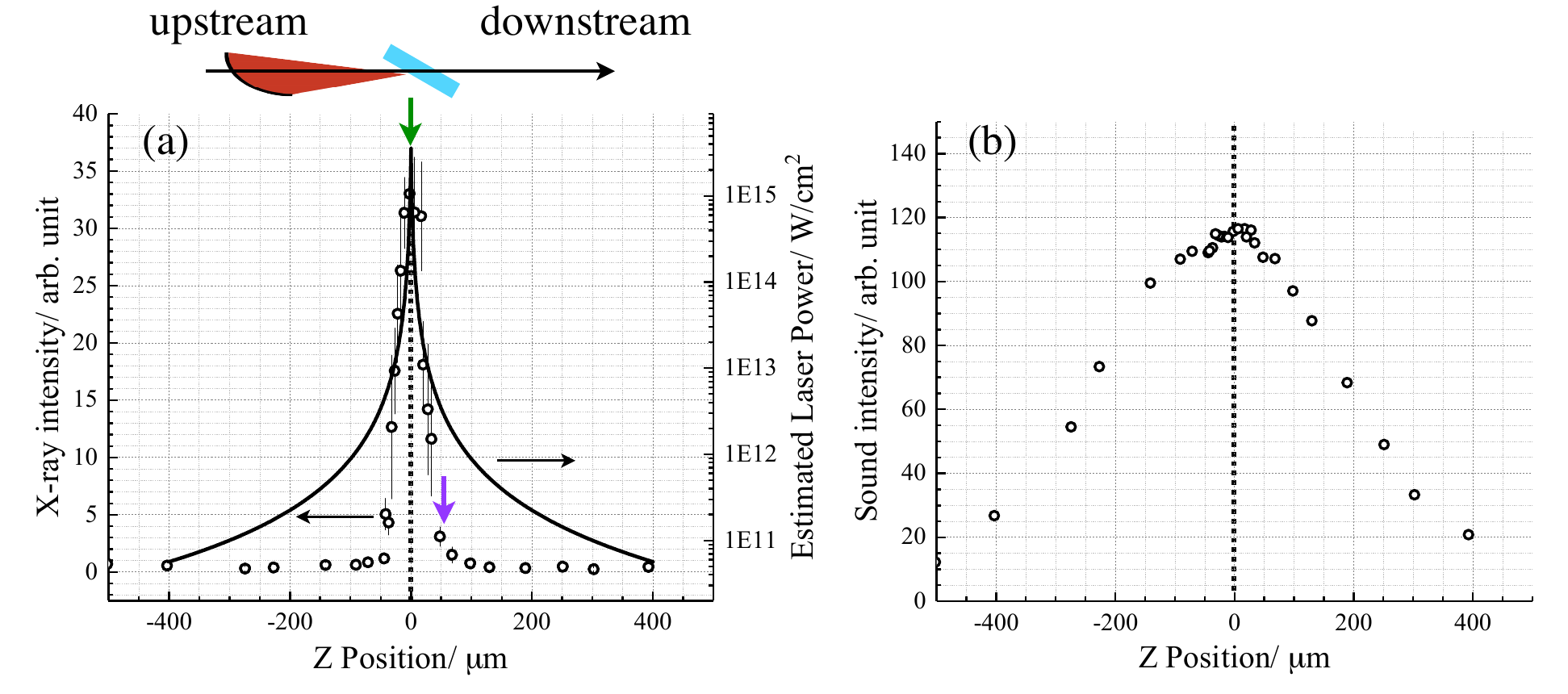}
 \caption{Emission intensity profiles of (a) X-ray and (b) sound along with the incident laser direction (Z-axis) under the single pulse excitation with the laser intensity (the main pulse) at 0.4 mJ/pulse. The position zero was set at the position of the water film when X-ray emission intensity was at the highest. The band width is estimated to be 37~$\mu$m for X-ray and 403~$\mu$m for sound, respectively.The estimated laser power along the Z-axis is also shown shown in (a). The downward arrows in green and violet in (a) indicate the Z-positions at Z=0 and Z=+50~$\mu$m of the water film for the pre-pulse spatial scan shown in Fig.4 (a) and (c), respectively. The details are in the text.} 
 \label{fgr:fig2}
\end{figure}
Figure 2 (a) and (b) show the emission intensity profile of X-ray and sound, respectively, as a function of the position of the water film along with the laser incident direction (Z-axis) when the laser intensity was set at 0.4 mJ/pulse. Effectively in these Z-scan experiments, the laser power on the water film continuously changes along the Z-axis as shown in Fig.4 (a). The bandwidth of the profile in X-ray is narrow at 37~$\mu$m, approximately twice as the thickness of water film, as reported previously~\cite{hatanaka2016, Hatanaka2018THz}. Though the sound emission shows a wider profile with the bandwidth at 403~$\mu$m, the peak position of the sound emission is the same as the peak position of X-ray emission. The sound detection in the experiments is time-integrated, therefore the measurements integrate the sound emitted through the whole duration of various ablation processes such as thermal decays and pressure changes related to heat emission, plasma formation, droplet formation, and shock-wave expansion~\cite{hatanaka2002}. However, such phenomena are surely enhanced under an effective coupling between the incident laser pulse and the solution surface. Then, fine positioning of the water film results in effective X-ray and sound emission. This result simply indicates that the sound emission intensity measured by a microphone with faster response time can be an indicator for X-ray emission when the water film is close to its best position for X-ray emission. In other words, sound detectors with much faster response time can be used for the fine positioning of water film for X-ray emission under the tight focusing condition. This also will contribute to combined usages of X-ray and sound for application in future. In earlier studies of light filamentation inside silica glass \cite{APL2008} or in air \cite{CPL2005} and their acoustic detection, it was observed that the entire length of the filament induced sound emission. This obscured the exact positioning of the sound source.

\begin{figure}[tb]
\centering
 \includegraphics[width=14cm]{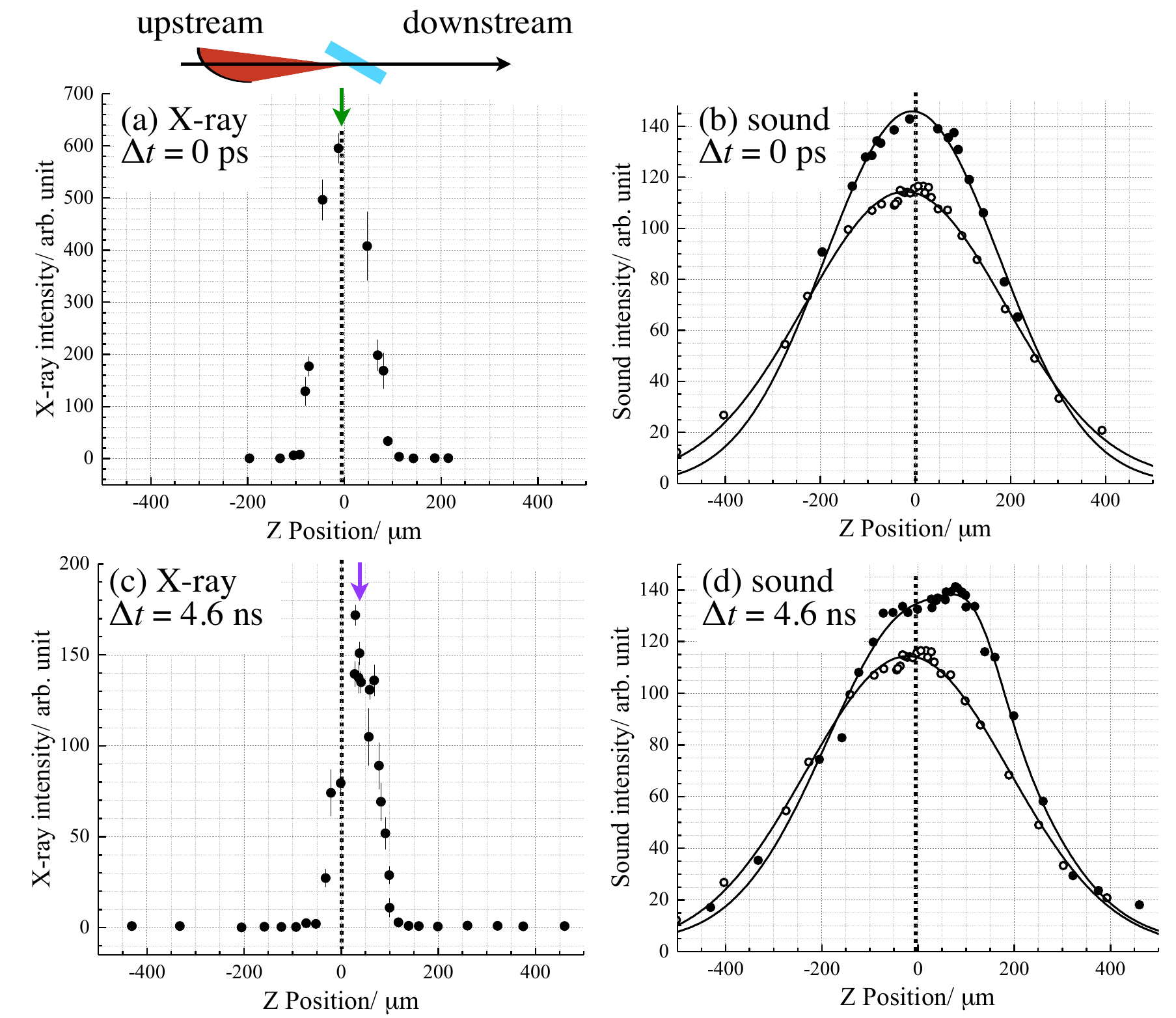}
 \caption{Emission intensity profiles of X-ray (a and c) and sound (b and d) under the double pulse excitation condition with the main pulse at 0.4 mJ/pulse (p-pol.) and a pre-pulse at 0.1 mJ/pulse (s-pol.). The delay time was set at 0 ps (a and b) and 4.6 ns (c and d). The open circles in (b) and (d) represent the result only with the main pulse irradiation shown in Fig.2(b). The downward arrows in green and violet in (a) and (c) indicate the Z-positions at Z=0 and Z= +50~$\mu$m of the water film for the pre-pulse spatial scan shown in Fig.4 (a and c), respectively. The details are in the text.  Solid lines are guides for readers.}
 \label{fgr:fig3}
\end{figure}
Figure 3 shows emission intensity profiles of X-ray (a and c) and sound (b and d) as a function of the position of the water film along the laser incident direction (Z-axis) under the double-pulse excitation conditions when the delay times are 0$\pm$1 ps (a and b) and 4.6$\pm$0.25 ns (c and d). The laser intensity is 0.4 mJ/pulse and 0.1 mJ/pulse for the main pulse and the pre-pulse, respectively. In this experiment, the main-pulse and the pre-pulse overlap spatially at the laser focus on the water film surface. As reported previously~\cite{17oe24109}, when the time delay is at 0 ps shown in Fig.~3(a), the X-ray emission shows a wider profile at 72~$\mu$m than the case of the single pulse excitation in Fig.~2(a). Under the same condition, the sound emission intensity as shown in Fig.~3(b) also increases but only at the positions close to zero with the same peak position. This is reasonable since the sound emission only from the pre-pulse irradiation is negligibly small and the sound intensity enhancement is due to the tight focusing on the water film surface. In the case of the delay time at 4.6~ns as shown in Fig. 3(c), when the initial processes of laser ablation such as transient surface formation and shock wave expansion are expected, asymmetric X-ray intensity enhancements are observed as reported previously~\cite{17oe24109}. Similar asymmetric intensity enhancements are also observed for the sound emission as shown in Fig. 3(d). As for dynamic morphological changes on solution surfaces induced by laser irradiations, there have been a lot of reports such as nanosecond near-IR laser irradiation on water flow \cite{Stasicki05}, 100-fs soft X-ray irradiation on water, ethanol, or protein suspension \cite{Wiedorn18}, and 300-fs UV laser irradiation on liquid toluene \cite{hatanaka98,hatanaka08}. On the basis of the knowledge on time-resolved surface scattering images \cite{hatanaka98}, at the delay time of 4.6 ns, the initial ablation process such as surface roughness or droplet formation has already started on the solution surface. This results in the additional peak position the water film along the Z-axis observed at 50~$\mu$m downstream from the best position of the water flow under the single pulse irradiation. The similar peak shift is also observed in the sound emission. These observations reflect that effective coupling between the incident laser pulses and the solution surface results in the enhanced emission of X-ray and sound and they correlate each other.

\begin{figure}[tb]
\centering
 \includegraphics[width=14cm]{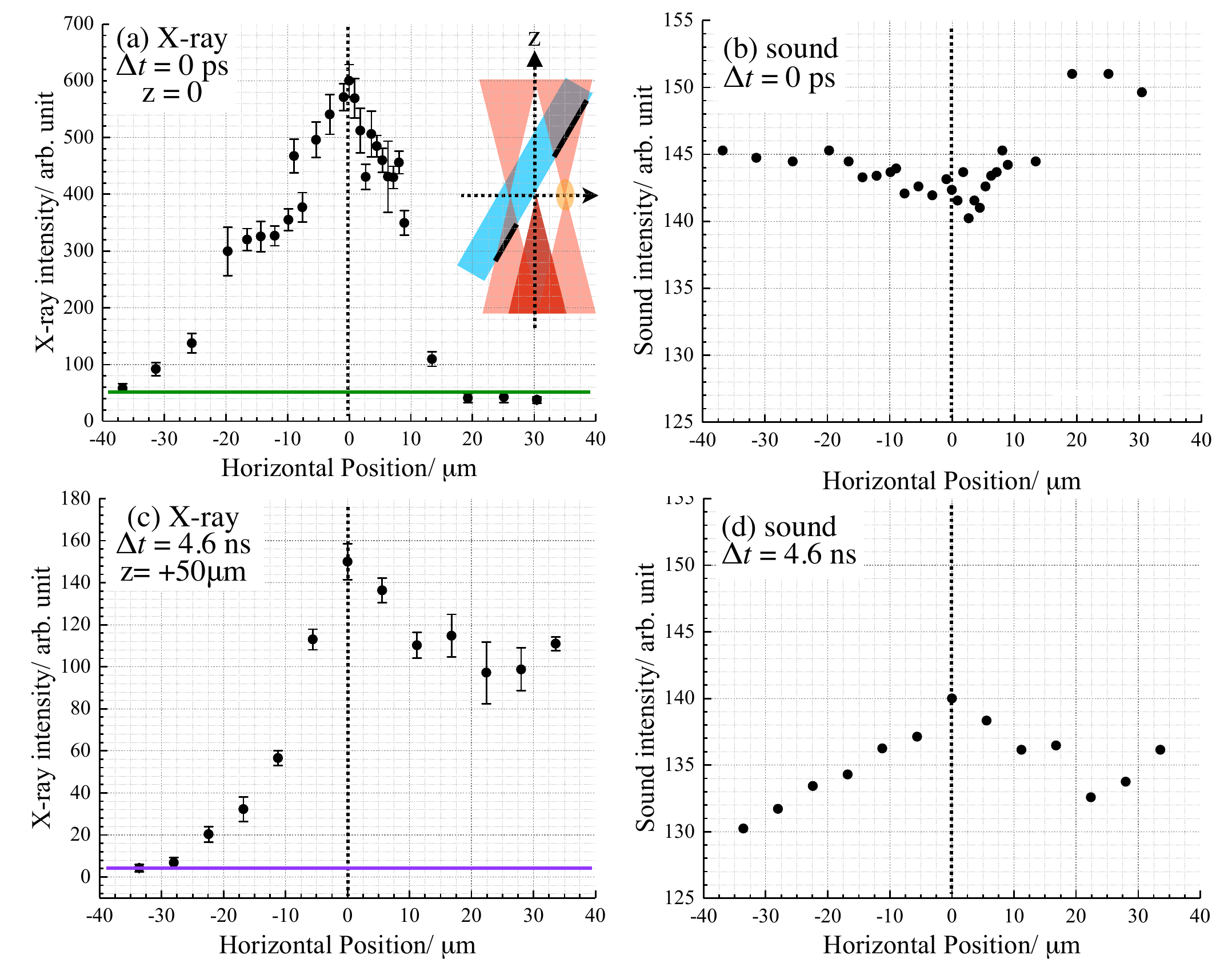}
 \caption{Emission intensity profiles of X-ray (a and c) and sound (b and d) as a function of the pre-pulse irradiation positions when the delay time at 0 ps (a and b) and 4.6 ns (c and d) and the water film position is Z=0 and Z= +50~$\mu$m as indicated by the arrows in Figs.2 and 3. The horizontal axis is defined as shown in Fig.1 or the inset in (a). The position zero indicates the perfect overlap between the main pulse and the pre-pulse. The horizontal lines in green (a) and violet (c) represent the X-ray intensity under the single pulse irradiation shown in Fig.2 (a). The insent in (b) shows a photograph of the plasma under the experimental condition.}
 \label{fgr:fig4}
\end{figure}
Figure 4 shows X-ray and sound emission under the same double pulse excitation conditions as in Fig.~3 but by scanning the pre-pulse irradiation positions on the solution surface horizontally. For practical applications in future of the synchronous X-ray and sound emission with higher intensity obtained under the double pulse irradiation, we believe that the knowledge on the spatial overlap of the pre-pulse with the main pulse is indispensable. In the case of the delay time at 4.6 ns, as shown in Fig. 4(c) and (d), the peak positions for the X-ray and the sound emission are the same with the asymmetric profile. The intensity decrease in X-ray and sound at the left side is due to the lower laser power per-unit-area of the pre-pulse at the solution surface as shown in the inset in Fig.~4(a). On the other hand, at the right side, the intensity of X-ray and sound is higher than that at the left side. The solution surface is behind the laser focus and the pre-pulse can be channeled effectively to the solution surface through self-generating pre-plasma. Interestingly, in the case of the double pulse excitation with the time delay at 0 ps, the sound intensity is smaller and the pre-pulse positioning dependence becomes flat when the pre-pulse and the main pulse overlap. The maximum emission of X-rays was observed at the zero position as shown in Fig.4 (a) and (b). Since the entire length of plasma is a sound source and was estimated to about 200~$\mu$m at longest as directly observed (see the inset of Fig.~4(b)), the dependence of the acoustic signal intensity on the pre-pulse irradiation position becomes flattened. Intensity clipping~\cite{CPL2005} in the fs-laser-induced filamentation by self-focusing defines the actual extent of the sound source. Under the double-pulse excitation, a plasma channel in air at the vicinity of the water film surface becomes elongated. It has been earlier observed that the acoustic signals of air breakdown by 150~fs laser pulses are strongly dependent on laser chirp~\cite{oyobuturi2010}. Localization of the breakdown in air in this case was obtained when the leading part of the laser pulse had a shorter wavelength while the trailing part was red-shifted. Filamentation in glass had smeared the acoustic signal due to due-localized plasma formation along the pulse propagation in glass~\cite{APL2008}. In this study a similar phenomenon was observed from fs-laser breakdown of a water film in air.

\section{Conclusions}

\label{sec4}
In this study, X-ray and sound emission from distilled water film when irradiated by tightly-focused femtosecond laser pulses in air were measured simultaneously under the single and double pulse excitation conditions. It has been clarified that such emission of transverse and longitudinal waves correlates each other and shows the same features and the peak positions when the water film changes its positions at the laser focus along the incident laser axis. Even under the double pulse excitation, as in the case of X-ray intensity enhancements reported previously, the sound intensity also increases. The pre-pulse spatial scan on the solution surface also shows a good correlation in the emission of X-ray and sound except the case of the double pulse excitation with the time delay at 0 ps. The sound intensity becomes independent on the focusing conditions due to filament formation. These findings are expected to contribute to future applications of combined emission of X-ray and sound as in the case of X-ray and THz wave synchronized emission \cite{Hatanaka2018THz} and further in the combination with THz wave and sound/ultrasound.

\section{Acknowdgments}
The author, KH, acknowledges MOST (107-2112-M-001-014-MY3) for its financial support.
\label{sec5}

\appendix





\bibliographystyle{elsarticle-num}

\bibliography{sample}

\end{document}